\documentclass[pdflatex,sn-chicago]{sn-jnl}
\usepackage{amsmath,amssymb,amsfonts}%
\usepackage{amsthm}
\usepackage{CJKutf8}
\usepackage{pgfplots}
\pgfplotsset{compat=1.18}
\usepackage{pdfpages}
\usepackage{url}        
\usepackage{ragged2e}   
\usepackage{tabularx} 
\usepackage{geometry}   
\usepackage{mathptmx}
\usepackage{mathrsfs}%
\usepackage[title]{appendix}%
\usepackage{xcolor}%
\usepackage{textcomp}%
\usepackage{manyfoot}%
\usepackage{algorithm}%
\usepackage{algorithmicx}%
\usepackage{algpseudocode}%
\usepackage{listings}%
\usepackage{hologo}
\usepackage{amsmath}
\usepackage{pgf-pie}
\usepackage[utf8]{inputenc}
\usepackage[T1]{fontenc}
\usepackage{subcaption}
\usepackage{float}
\usepackage{tcolorbox}%
\usepackage{longtable}
\usepackage{tikz}
\usepackage{amsmath}
\usepackage{xcolor}
\usepackage{rotating}
\usepackage{siunitx}
\usepackage{graphicx}
\DeclareUnicodeCharacter{CD5C}{Choi} 
\DeclareUnicodeCharacter{D76C}{Hee}  
\DeclareUnicodeCharacter{BCF5}{Bok}  
\usepackage{etaremune}
\usetikzlibrary{shapes,arrows,calc,positioning,shadows.blur,shapes.geometric,decorations.pathreplacing,decorations.markings,patterns,backgrounds,fit}
\usepackage{mdframed}
\usepackage{amsmath, amssymb, bm, physics}
\usepackage{tabularx} 
\usepackage{hyperref} 
\hypersetup{
    colorlinks   = true, 
    urlcolor     = blue, 
    linkcolor    = blue, 
    citecolor    = blue, 
}
\usepackage{array}
\usepackage{multicol}
\usepackage{enumitem}
\usepackage{xcolor}
\usepackage{booktabs}
\newcommand{\mycline}[1]{%
    \noalign{\hrule height 0.4pt} 
    \noalign{\vskip -0.4pt} 
    \hline
    \noalign{\vskip -0.4pt} 
}
\definecolor{lightbluea}{RGB}{176,224,230}
\definecolor{MintCream}{RGB}{245, 255, 250}
\definecolor{PeachPuff}{RGB}{255, 218, 185}

\begin{document}

\title[Real-Time Scream Detection and Position Estimation for Worker
Safety in Construction Sites]{Real-Time Scream Detection and Position Estimation for Worker
Safety in Construction Sites}  

\author[1]{\fnm{Bikalpa} \sur{Gautam}}
\author*[2]{\fnm{Anmol} \sur{Guragain}}
\author[3]{\fnm{Sarthak} \sur{Giri}}

\affil[1]{\orgdiv{Research Officer}, \orgname{Singapore Institute of Manufacturing Technology}, \orgaddress{\country{Singapore}}}
\affil[2]{\orgdiv{Research Assistant}, \orgname{Nepal Applied Mathematics and Informatics Institute for Research (NAAMII)}, \orgaddress{\country{Nepal}}}
\affil[3]{\orgdiv{Software Engineer}, \orgname{ION Group}, \orgaddress{\country{India}}}

\abstract{The construction industry faces high risks due to frequent accidents, often leaving workers in perilous situations where rapid response is critical. Traditional safety monitoring methods, including wearable sensors and GPS, often fail under obstructive or indoor conditions. This research introduces a novel real-time scream detection and localization system tailored for construction sites, especially in low-resource environments. Integrating Wav2Vec2 and Enhanced ConvNet models for accurate scream detection, coupled with the GCC-PHAT algorithm for robust time delay estimation under reverberant conditions, followed by a gradient descent-based approach to achieve precise position estimation in noisy environments. Our approach combines these concepts to achieve high detection accuracy and rapid localization, thereby minimizing false alarms and optimizing emergency response. Preliminary results demonstrate that the system not only accurately detects distress calls amidst construction noise but also reliably identifies the caller's location. This solution represents a substantial improvement in worker safety, with the potential for widespread application across high-risk occupational environments. The scripts used for training, evaluation of scream detection, position estimation, and integrated framework will be released at \hyperlink{https://github.com/Anmol2059/construction_safety}{\url{https://github.com/Anmol2059/construction_safety}}.}



\renewcommand{\thefootnote}{\textit{*}}
\footnotetext{Corresponding Author: \href{mailto:anmol.guragain@naamii.org.np}{anmol.guragain@naamii.org.np}}

\renewcommand{\thefootnote}{\textit{ab. \& sym.}} 
\footnotetext{The following abbreviations and symbols are used in this manuscript:\\
 \begin{tabular}{@{}ll@{\hskip 1cm}ll@{\hskip 1cm}ll@{}}
 \textbf{Abbreviations} & & \textbf{Symbols} & & \textbf{Symbols} & \\
 PPE & Personal Protective Equipment & $R$ & Sample rate (Hz) & $ReLU$ & Rectified Linear Unit activation function\\
 GPS & Global Positioning System & $T$ & Duration (seconds) & $W_{ih}$, $W_{hh}$ & Weight matrices for LSTM\\
 TDOA & Time Difference of Arrival & $D$ & Fixed duration (seconds) & $b_{ih}$, $b_{hh}$ & Bias terms in LSTM\\
 Wav2Vec2 & Waveform-to-Vector-2 & $MFCC_{t,n}$ & Mel-frequency cepstral coefficient & $x_t$ & Input at time $t$\\
 DNN & Deep Neural Network & $X_{t,f}$ & Short-time Fourier transform & $h_t$ & Hidden state at time $t$\\
 SVC & Support Vector Classifier & $F$ & Number of filters & $\nabla Loss$ & Gradient of the loss function\\
 LR & Logistic Regression & $Acc$ & Accuracy & $\eta$ & Learning rate\\
 RF & Random Forest & $P$ & Precision & $c$ & Speed of sound\\
 KNN & K-Nearest Neighbors & $R$ & Recall & $d_i$, $d_j$ & Distances to mics $i$ and $j$\\
 DT & Decision Tree & $F1$ & F1-score & $Loss$ & Loss function for position estimation\\
 MLP & Multi-Layer Perceptron & $\tau$ & Threshold (ROC or TDOA) & $\Psi_{PHAT}(f)$ & PHAT function\\
 XGB & XGBoost & $B$ & Batch size & $R_{PHAT}(\tau)$ & GCC-PHAT cross-correlation\\
 LGBM & LightGBM & $T$ & Sequence length & $p_{new}$, $p_{old}$ & New and old position estimates\\
 CAT & CatBoost & $F$ & Feature dimension & $\partial Loss / \partial x$ & Partial derivative of loss\\
 MFCC & Mel-frequency Cepstral Coefficients & $Conv1D(X)$ & 1D convolution of $X$ & $\hat{P}(q)$ & SRP-PHAT power function\\
 STFT & Short-Time Fourier Transform & $MaxPool$ & Max-pooling operation & $\rho_{ij}(\tau_i, \tau_j)$ & Pairwise correlation factor\\
 SNR & Signal-to-Noise Ratio & $EER$ & Equal Error Rate & $tr(R)$ & Trace of matrix $R$\\
 GCC-PHAT & Generalized Cross-Correlation & $R_{x_i,y_j}(\tau)$ & Cross-correlation function & $h_{DW}(\tau)$ & MCCC weighting function \\
 SRP-PHAT & Steered Response Power with PHAT & & & & \\
 \end{tabular}
}
\maketitle
\vspace{-25pt}
\section{Introduction}\label{sec1}

The construction industry has been viewed as particularly dangerous due to its high rates of workplace injuries and fatalities \cite{Almaskati2024, RAHEEM2014276}. Falls, entrapments, and collapsing structures can leave workers in dire situations where timely rescue is critical. Ensuring the safety of construction projects and mitigating financial and human losses resulting from accidents are issues confronting the global construction industry \cite{Chan2021}. While safety protocols have evolved to reduce such risks, traditional methods for locating workers in distress remain inadequate, particularly when the injured or trapped individual cannot be visually identified. The ability to vocally signal for help—such as through screaming—presents a natural means for workers to draw attention to their predicament, yet the challenge lies in effectively detecting these sounds amidst the cacophony of a construction site and accurately pinpointing the location of the person in need.

Historically, efforts to improve worker safety have focused on enhancing personal protective equipment (PPE) \cite{Atasoy2024, AMMAD20213495, Sehsah2020ppe} and developing better monitoring systems through wearable devices, such as sensors and GPS trackers \cite{idris2023smart, devi2023safety, angelia2021wireless, he2018application, choi2017wearable, mohanapriya2022helmet, sreenivasaraja2023vr, rashid2017trajectory, kim2020iot, wang2016low, jang2013safety, arcayena2019arduino}. These technologies, however, have limitations. GPS systems often struggle in indoor \cite{RAZAVI2012128} or obstructed environments, and wearable sensors can be damaged or rendered ineffective during accidents \cite{Boniphace2024, Choi2020}. Acoustic detection methods, which leverage the human voice as a distress signal, offer a compelling alternative, but previous approaches have been hampered by environmental noise, the complexity of sound localization, and issues of accuracy in determining the precise position of the signal's source.

Past attempts at using sound-based systems for safety have primarily focused on detecting general noise patterns rather than human screams specifically \cite{LEE2020103127, YANG2023106093}. Elelu et al. \cite{Elelu2023} introduced a novel audio-based machine learning model that enhances the detection of collision hazards on construction sites by improving auditory situational awareness, even in environments with loud ambient noise. These systems often rely on basic audio thresholding techniques, where any loud or unusual sound triggers an alert. While this can be useful in certain settings, it leads to numerous false positives—alarms triggered by construction machinery or background noise rather than actual distress signals. Moreover, previous methods for sound localization, such as basic triangulation techniques, suffer from inaccuracies, especially in environments where sound waves are obstructed by materials or distorted by reverberation.

Recent advances in machine learning and signal processing have opened new possibilities for more sophisticated audio-based detection systems. Models that can analyze and recognize specific sound signatures—such as human distress signals—have shown promise in overcoming the noise problem. Self-supervised learning (SSL) models like Wav2Vec2 \cite{Baevski2020} and WavLM \cite{Chen2022} have demonstrated state-of-the-art results in audio feature extraction and across various downstream tasks. Not limited to scream detection, these SSL models have also excelled in major audio classification tasks with significant noise, such as singing voice deepfake detection \cite{Guragain2024}, speech emotion recognition \cite{siriwardhana2020jointly}, and automatic speaker verification \cite{yadav2022survey}. Given their success in extracting robust audio features under noisy conditions, we decided to leverage SSL models in our framework.

At the same time, developments in localization techniques, particularly Time Difference of Arrival (TDOA) methods, offer greater accuracy in pinpointing the source of a detected sound. These innovations have laid the groundwork for a new generation of safety systems capable of not only detecting distress calls but also determining their precise origin in real time.

In this research, we build on these advancements by introducing a novel system that integrates state-of-the-art machine learning for screaming detection with TDOA-based localization using the GCC-PHAT method. Our approach addresses the shortcomings of past methods, offering a highly accurate and reliable means of locating workers in distress, even in the chaotic environment of a construction site. In the following sections, we will discuss the methodology and the technical details behind the system, demonstrating its potential for real-world applications in enhancing worker safety.

\section{Proposed Framework}
\vspace{-15pt}
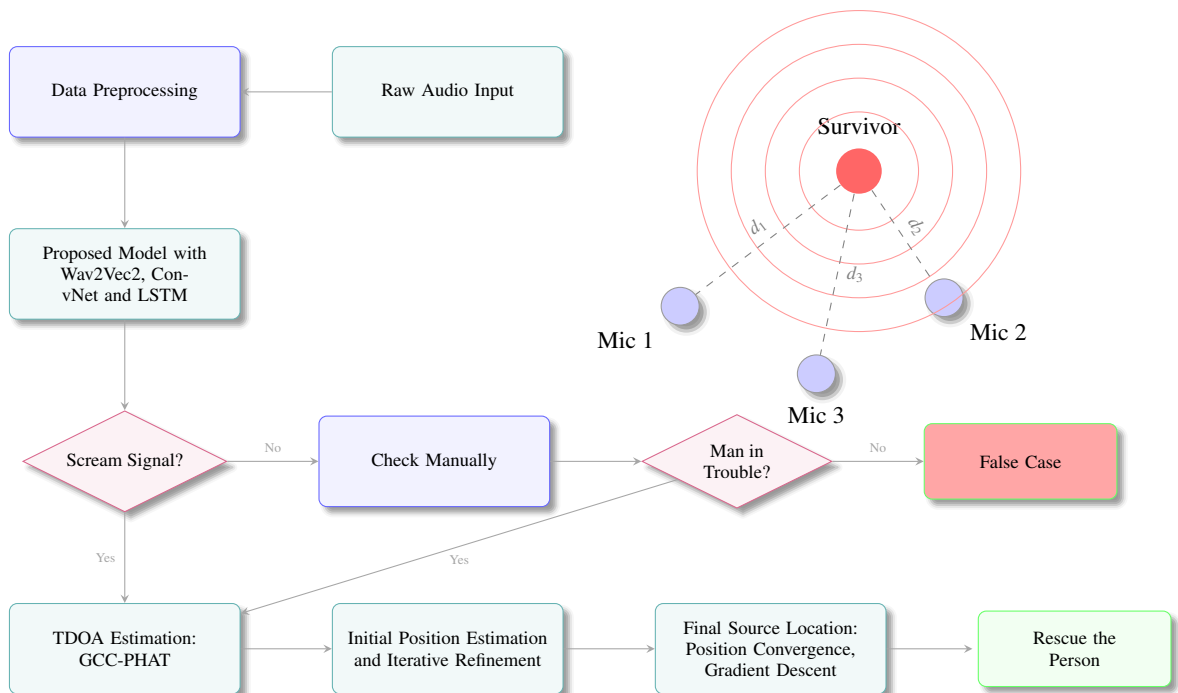
\begin{figure}[h]
\begin{center}
\begin{tikzpicture}[
scale=0.7,
node distance=1.2cm,
process/.style={
rectangle,
minimum width=3cm,
minimum height=1.2cm,
text width=2.8cm,
align=center,
draw=blue!60,
fill=blue!5,
rounded corners=3pt,
blur shadow={shadow blur steps=3},
font=\footnotesize
},
decision/.style={
diamond,
aspect=2,
draw=purple!60,
fill=purple!5,
minimum width=2.5cm,
align=center,
blur shadow={shadow blur steps=3},
font=\footnotesize
},
sensor/.style={
circle,
draw=gray!80,
fill=blue!20,
minimum size=0.5cm,
blur shadow={shadow blur steps=2}
},
algorithm/.style={
rectangle,
minimum width=3cm,
minimum height=1.2cm,
text width=2.8cm,
align=center,
draw=teal!60,
fill=teal!5,
rounded corners=3pt,
blur shadow={shadow blur steps=3},
font=\footnotesize
},
result/.style={
rectangle,
minimum width=2.5cm,
minimum height=1cm,
text width=2.3cm,
align=center,
draw=green!60,
fill=green!5,
rounded corners=3pt,
blur shadow={shadow blur steps=3},
font=\footnotesize
},
arrow/.style={
->,
thin,
>=stealth,
gray!70
}
]
\node[process] (input) {Data Preprocessing};
\node[algorithm, right=of input] (preprocess) {Raw Audio Input};
\node[algorithm, below=of input] (model) {Proposed Model with Wav2Vec2, ConvNet and LSTM};
\node[decision, below=of model] (decide) {Scream Signal?};
\node[process, right=of decide] (check) {Check Manually};
\node[decision, right=of check] (trouble) {Man in\\Trouble?};
\node[result, fill=red!35, right=of trouble] (false) {False Case};

\node[algorithm, below=of decide] (tdoa) {TDOA Estimation:\\GCC-PHAT};
\node[algorithm, right=of tdoa] (snr) {Initial Position Estimation and Iterative Refinement};
\node[algorithm, right=of snr] (position) { Final Source Location:\\ Position Convergence, Gradient Descent};
\node[result, right=of position] (response) {Rescue the\\Person};

\begin{scope}[xshift=13cm, yshift=-1.5cm, scale=1.6]
    \node[sensor] (mic1) at (-1.6,-1.6) {};
    \node[below left=0.05cm of mic1, font=\small] {Mic 1};
    \node[sensor] (mic2) at (1.5,-1.5) {};
    \node[below right=0.05cm of mic2, font=\small] {Mic 2};
    \node[sensor] (mic3) at (0,-2.4) {};
    \node[below=0.05cm of mic3, font=\small] {Mic 3};
    \node[circle, fill=red!60, minimum size=0.6cm] (survivor) at (0.5,0) {};
    \node[above=0.05cm of survivor, font=\small] {Survivor};
    \foreach \r in {0.7,1.1,1.5,1.9}{
        \draw[red!40] (survivor) circle (\r cm);
    }
    \draw[dashed, gray] (survivor) -- (mic1) node[midway, above, sloped, font=\footnotesize] {$d_1$};
    \draw[dashed, gray] (survivor) -- (mic2) node[midway, above, sloped, font=\footnotesize] {$d_2$};
    \draw[dashed, gray] (survivor) -- (mic3) node[midway, right, font=\footnotesize] {$d_3$};
\end{scope}

\draw[arrow] (preprocess) -- (input);
\draw[arrow] (input) -- (model);
\draw[arrow] (model) -- (decide);
\draw[arrow] (decide) -- node[above, font=\tiny] {No} (check);
\draw[arrow] (check) -- (trouble);
\draw[arrow] (trouble) -- node[above, font=\tiny] {No} (false);
\draw[arrow, out=270, in=180] (trouble) -- node[below, font=\tiny] {Yes} (tdoa);
\draw[arrow] (decide) -- node[left, font=\tiny] {Yes} (tdoa);
\draw[arrow] (tdoa) -- (snr);
\draw[arrow] (snr) -- (position);
\draw[arrow, shorten >=2pt] (position) -- (response);
\end{tikzpicture}
\end{center}
\caption{Illustration of the proposed system's workflow, from audio input and signal processing to detection, localization, and emergency response.}
\label{fig:emergency-system-workflow}
\end{figure}
\vspace{-10pt}
The proposed framework of our research is given by Fig. \ref{fig:emergency-system-workflow}. We have designed it to significantly improve worker safety in construction environments by detecting distress signals, such as screams, and accurately determining the location of the person in distress. The system begins with a network of strategically positioned high-fidelity microphones across the construction site. These microphones capture the ambient sounds in real time, monitoring the environment continuously for any signs of danger.

To ensure uniformity and compatibility with the Wav2Vec2 model, each audio sample is captured in 10-second segments. After every 10 seconds, the audio is sent for model inference, creating a loop of continuous monitoring. Each audio segment undergoes preprocessing where it is padded or truncated to exactly 10 seconds, ensuring consistency. Additionally, the sampling rate (SR) is standardized to 16 kHz to align with the input requirements of Wav2Vec2 and to maintain a uniform standard across diverse microphone types, as they may vary in capture specifications.

\begin{figure}[H]
    \centering
    \includegraphics[width=\linewidth,height=0.15\textheight]{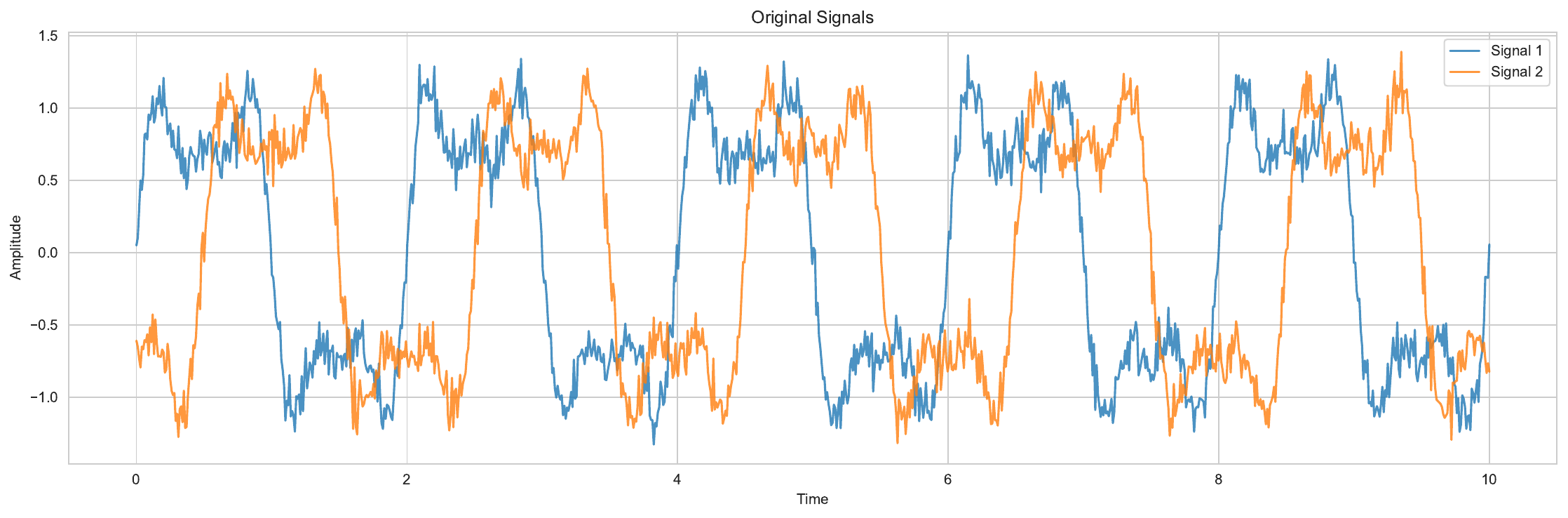}
    \caption{Sample signals received by the sensors.}
    \label{fig:original-signals}
\end{figure}

\begin{figure}[H]
    \centering
    \includegraphics[width=\linewidth,height=0.15\textheight]{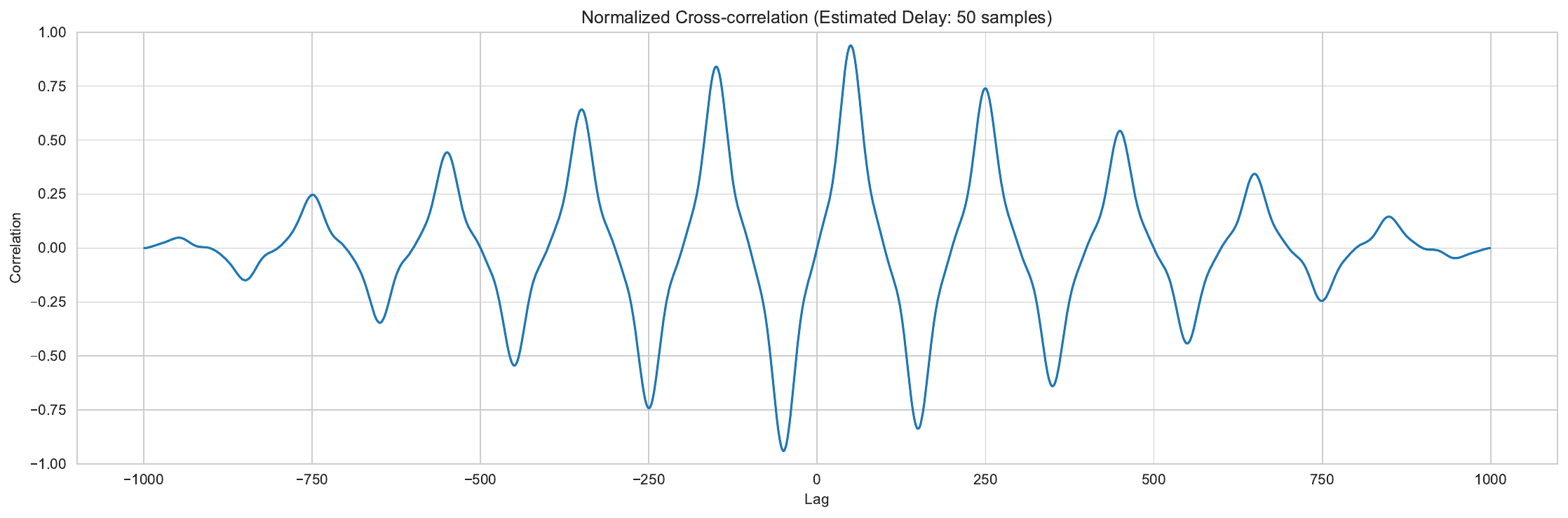}
    \caption{The corresponding correlation graph for Time Difference of Arrival (TDOA) using GCC-PHAT.}
    \label{fig:full-correlation}
\end{figure}

\begin{figure}[H]
    \centering
    \includegraphics[width=\linewidth,height=0.15\textheight]{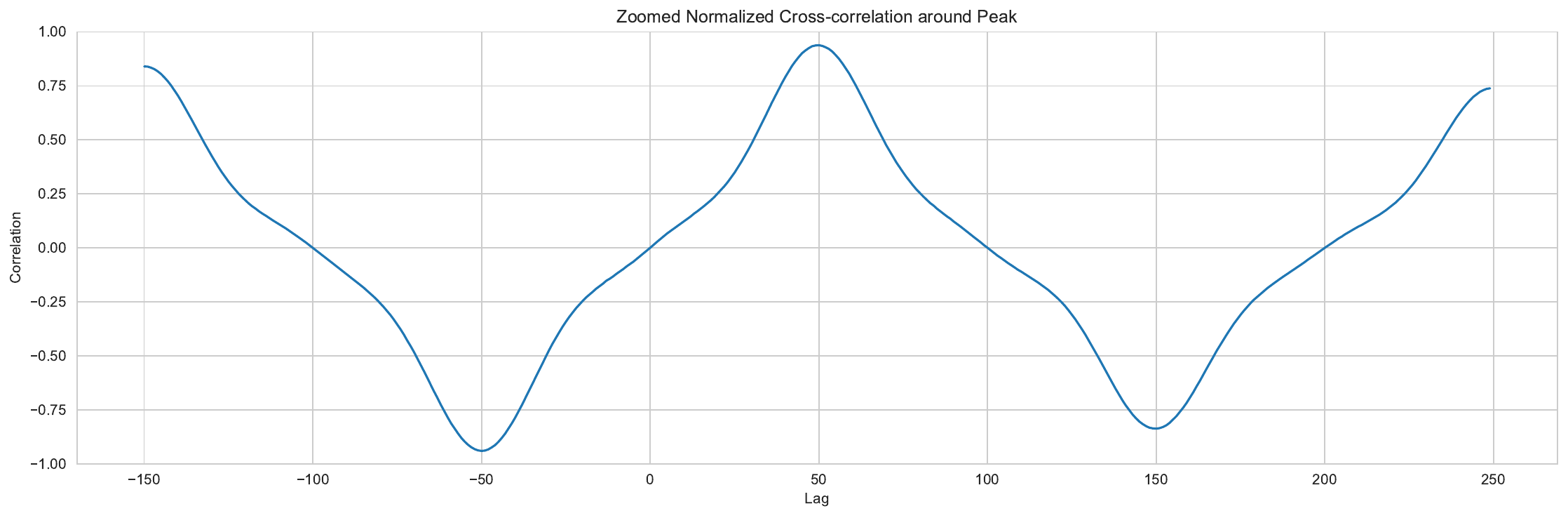}
    \caption{Zoomed-in view of the correlation graph for TDOA using GCC-PHAT, highlighting finer details of the correlation peaks.}
    \label{fig:zoomed-correlation}
\end{figure}

\begin{figure}[H]
    \centering
    \includegraphics[width=\linewidth,height=0.20\textheight]{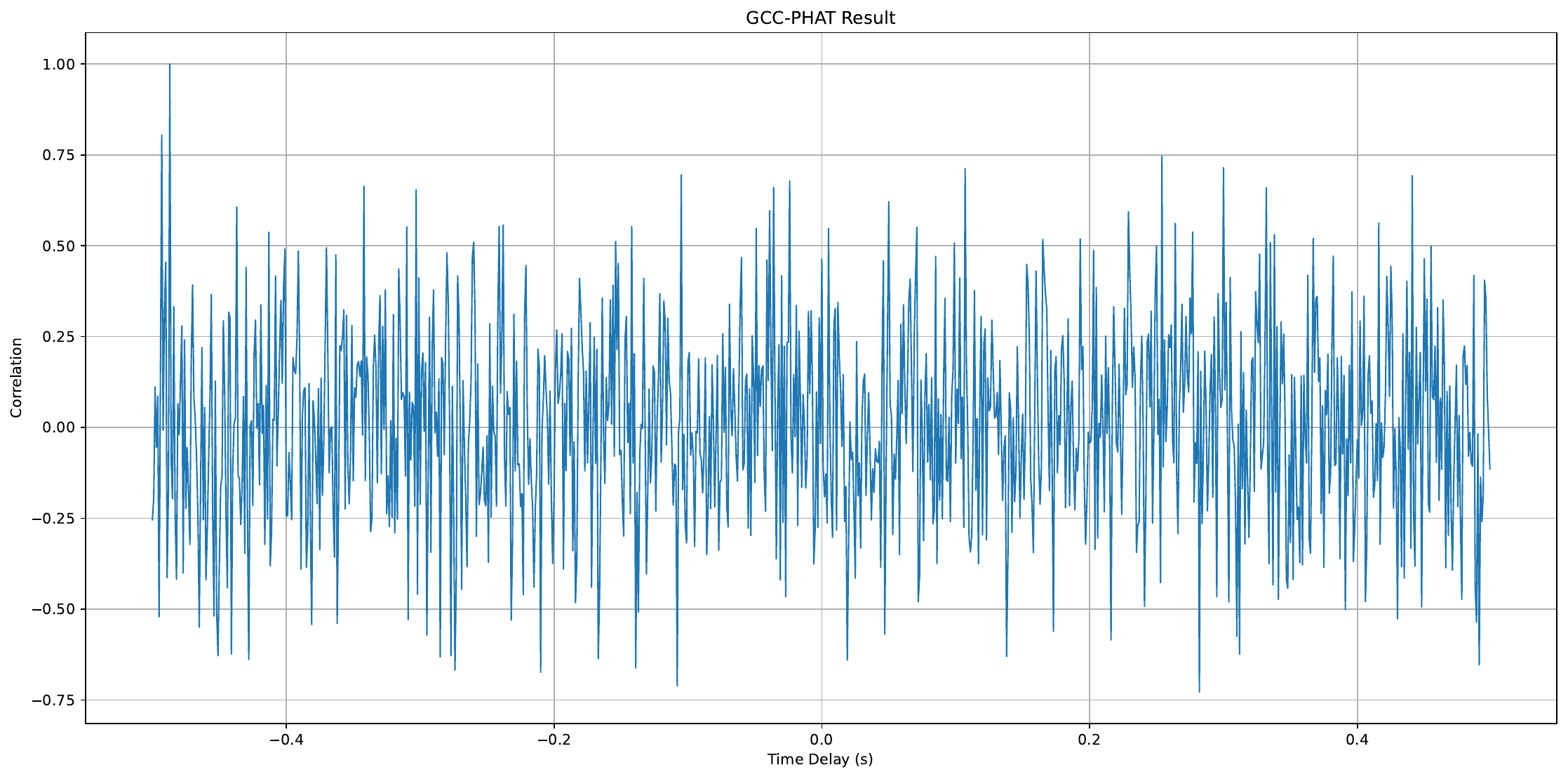}
\caption{GCC-PHAT result showing high-density correlation data over a short time interval.}
    \label{fig:gcc-phat}
\end{figure}

The audio is then segmented into frames and analyzed using a state-of-the-art Wav2Vec2 model, which extracts detailed acoustic features from the signals. These features are subsequently fed into an enhanced convolutional neural network, followed by an LSTM architecture and classifier to distinguish distress sounds from typical construction noises. Once a scream is detected, the system initiates the localization process by calculating the Time Difference of Arrival (TDOA) between strategically positioned microphones across the construction site. Using the GCC-PHAT algorithm, which emphasizes phase information to handle reverberations, the system identifies the time delays at which the scream arrives at each microphone shown in Fig. \ref{fig:original-signals}, \ref{fig:full-correlation}, \ref{fig:zoomed-correlation}, and \ref{fig:gcc-phat}. This enables the calculation of the TDOA values for microphone pairs, which are crucial for determining the source location. The next step involves scientifically relating these observed TDOA values to theoretical models of sound propagation: for each microphone pair, the TDOA is compared to the theoretical time difference based on the known distance between the microphones and the speed of sound in air.

\begin{figure}[h]
    \centering
    \includegraphics[width=0.4\textwidth]{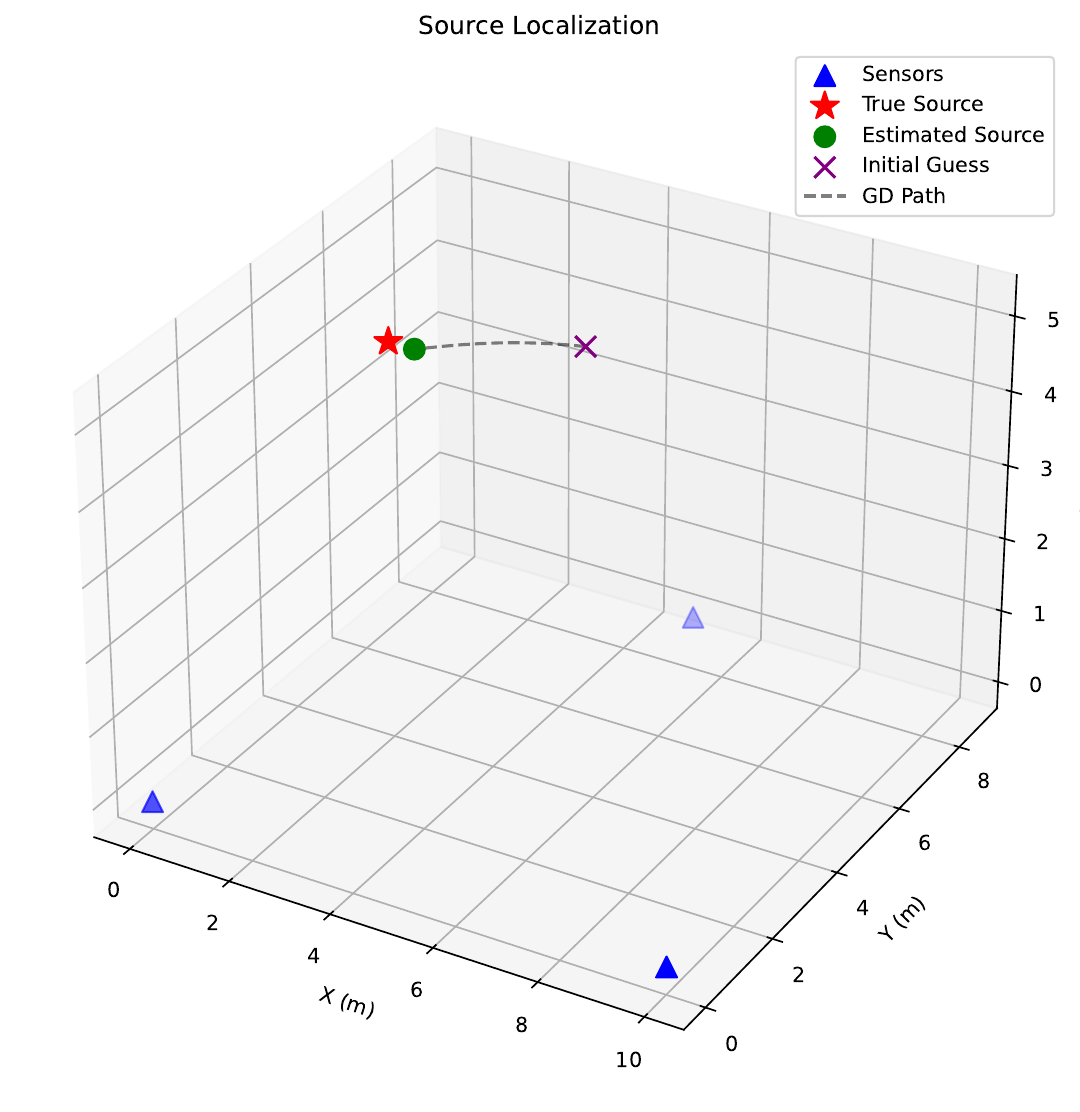}
    \caption{3D visualization of source localization with sensor positions, true source position, and estimated source position using gradient descent. The dashed line represents the gradient descent path converging toward the estimated position.}
    \label{gradientdescentpath}
\end{figure}

This discrepancy between observed TDOA and theoretical time delay is minimized using a gradient descent algorithm, which iteratively adjusts the estimated position to minimize the descrepency, as shown in Fig. \ref{gradientdescentpath}. By minimizing this discrepancy, the gradient descent method accommodates real-world nonlinearities and improves positional accuracy, particularly in noisy or reflective environments. This continuous refinement provides the system with a highly accurate estimate of the distressed worker's coordinates, enabling real-time emergency response. The localization component, thus, ensures that the system not only detects distress signals but also offers precise intervention capabilities, significantly enhancing safety and response effectiveness on construction sites. If an automated detection system fails to identify a scream, but screams are audibly perceived through manual observation, a manual verification process is initiated. Following this verification, the established Time Difference of Arrival (TDOA) calculation and localization procedures, as outlined above, are applied to pinpoint the sound source accurately.

\section{Experimental Setup: Scream Detection}
This study investigates scream detection by comparing the performance of various traditional machine learning classifiers with a deep neural network (DNN) model based on Wav2Vec2 and enhanced ConvNet architectures. Our method includes thorough preprocessing, feature extraction, and training with distinct evaluation metrics.

\renewcommand{\thefootnote}{\textit{a}}

\subsection{Dataset}
The dataset used in this study is a combination of two publicly available datasets from Kaggle.\footnote{Data sources: \href{https://www.kaggle.com/datasets/aananehsansiam/audio-dataset-of-scream-and-non-scream}{aananehsansiam/audio-dataset-of-scream-and-non-scream}\\ \href{https://www.kaggle.com/datasets/whats2000/human-screaming-detection-dataset}{whats2000/human-screaming-detection-dataset}.}


\begin{figure}[htb]
    \centering
    \begin{tikzpicture}
        \pie[
            radius=2,
            color={MintCream, PeachPuff}
        ]{
            37/Scream Files,
            63/Non-Scream Files
        }
    \end{tikzpicture}
    \caption{Overall distribution of Scream vs Non-Scream files (Total: 6,621 files).}
    \label{piechart}
\end{figure}
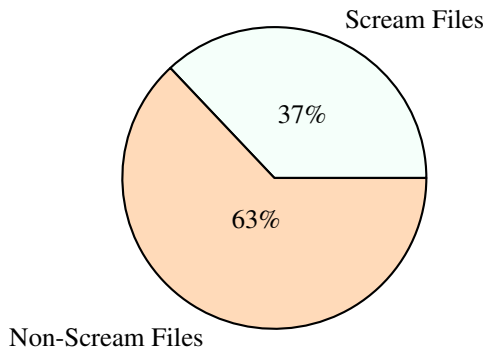

The dataset comprises a total of 6,621 audio files, with 2,445 classified as scream files, representing 37\% of the dataset, and the remaining 4,176 files, or 63\%, as non-scream files (see Fig. \ref{piechart} and \ref{Number of files}). The average length of these recordings is 6.41 seconds (see Fig. \ref{duration} for full details), with the top 10\% of files extending up to 10 seconds. Sampling rates are split between two standards, 16,000 Hz and 44,100 Hz, ensuring high-quality capture across different audio settings.

\begin{figure}[htb]
    \centering
    \begin{minipage}{0.49\textwidth}
        \centering
        \begin{tikzpicture}
            \begin{axis}[
                title=Comparison of file distribution across datasets,
                ybar,
                bar width=20pt,
                enlarge x limits=0.7,
                width=1\textwidth,
                height=0.8\textwidth,
                symbolic x coords={aanahensiam, whats2000},
                xtick=data,
                ylabel=Number of Files,
                legend style={at={(0.5,-0.2)}, anchor=north},
                legend columns=2,
                nodes near coords,
                xlabel={Dataset (Sampling Rate)},
                xticklabel style={align=center},
                xticklabels={aanahensiam\\(16000Hz), whats2000\\(44100Hz)}
            ]
                \node[anchor=north west] at (rel axis cs:0.02,0.98) {\small Total Non-Scream files: 4,176};
                \node[anchor=north west] at (rel axis cs:0.02,0.93) {\small Total Scream files: 2,445};
                \addplot coordinates {(aanahensiam,1545) (whats2000,2631)};
                \addplot coordinates {(aanahensiam,1583) (whats2000,862)};
                \legend{Non-Scream Files, Scream Files}
            \end{axis}
        \end{tikzpicture}
        \caption{Comparison of file distribution across datasets.}
        \label{Number of files}
    \end{minipage}%
    \hfill
    \begin{minipage}{0.48\textwidth}
        \centering
        \begin{tikzpicture}
        \begin{axis}[
            width=\textwidth,
            height=0.8\textwidth,
            xlabel=Duration (seconds),
            ylabel=Number of Files,
            title=Distribution of Audio Durations,
            ybar=0pt,
            bar width=1,
            xtick={0,...,11},
            xticklabels={0-1,1-2,2-3,3-4,4-5,5-6,6-7,7-8,8-9,9-10,10-11,11-12},
            xtick align=inside,
            x tick label style={rotate=45,anchor=east},
            ymajorgrids=true,
            grid style=dashed,
            legend style={at={(0.02,0.98)}, anchor=north west},
            legend cell align={left}
        ]
        \addplot[fill=blue!20, draw=black] coordinates {
            (0,113)
            (1,1495)
            (2,595)
            (3,486)
            (4,189)
            (5,288)
            (6,58)
            (7,23)
            (8,24)
            (9,212)
            (10,3195)
            (11,3)
        };
        \draw[red, dashed, thick] (axis cs:6.41,0) -- (axis cs:6.41,3500) 
            node[pos=0.98, anchor=east] {\fontsize{7pt}{9pt}\selectfont Mean Duration: 6.41 sec};
        \draw[green!50!black, dashed, thick] (axis cs:10.00,0) -- (axis cs:10.00,3500)
            node[pos=0.938, anchor=east] {\fontsize{7pt}{9pt}\selectfont 90th Percentile Duration: 10.00 sec};
        \end{axis}
        \end{tikzpicture}
        \caption{Distribution of audio durations.}
        \label{duration}
    \end{minipage}
\end{figure}

\subsection{Traditional Machine Learning Models}

Previous studies on scream detection in audio have shown that various traditional machine learning models perform well in scream detection tasks \cite{saeed2021initial, nandwana2015robust, potharaju2019classification}.
We lacked a standard benchmark for our combined dataset; thus, we also experimented with a variety of traditional classifiers to establish a baseline for performance. The classifiers were sourced from the standard Scikit-learn library or their respective libraries, providing reliable implementations for audio event detection tasks.

\subsubsection{Data Preprocessing}
All audio samples were standardized to a sample rate of \( R = 16000 \, \text{Hz} \) and trimmed or padded to a fixed duration of \( T = R \times D \) samples, where \( D = 5 \, \text{seconds} \). To capture meaningful audio features, Mel-frequency cepstral coefficients (MFCCs) were computed for each audio segment. The MFCCs, commonly used in audio classification, represent the power spectrum of audio samples, defined as:
\[
\text{MFCC}_{t, n} = \log \left( \sum_{f=1}^{F} |X_{t, f}| \cdot \cos\left(\frac{\pi n}{F}\right) \right)
\]
where \( X_{t, f} \) denotes the short-time Fourier transform (STFT) of the audio at time \( t \) and frequency \( f \), with \( F \) representing the number of filters.

\subsubsection{Classifiers}
We utilized a range of traditional classifiers, each offering unique algorithmic advantages for detecting scream events in audio. The classifiers are as follows:

\renewcommand{\thefootnote}{\textit{b}}

\begin{itemize}
    \item \textbf{Support Vector Classifier (SVC):} A linear SVC, which creates a hyperplane to maximize the margin between two classes\footnote{Implemented from Scikit-learn’s \texttt{SVC}: \url{https://scikit-learn.org/stable/modules/svm.html}}. The linear kernel offers a balance between computational efficiency and classification power in binary scream detection.

    \renewcommand{\thefootnote}{\textit{c}}
    
    \item \textbf{Logistic Regression (LR):} LR models t    he probability of binary outcomes using the sigmoid function\footnote{Implemented from Scikit-learn’s \texttt{LogisticRegression}: \url{https://scikit-learn.org/stable/modules/linear_model.html\#logistic-regression}}. By mapping input features to probabilities, LR provides interpretable likelihoods for scream events.

    \renewcommand{\thefootnote}{\textit{d}}

    \item \textbf{RandomForestClassifier:} An ensemble of decision trees that improves predictive performance by averaging over multiple trees\footnote{Implemented from Scikit-learn’s \texttt{RandomForestClassifier}: \url{https://scikit-learn.org/stable/modules/generated/sklearn.ensemble.RandomForestClassifier.html}}. This model handles feature interactions effectively, enhancing performance in diverse audio contexts.

    \renewcommand{\thefootnote}{\textit{e}}

    \item \textbf{K-Nearest Neighbors (KNN):} This classifier predicts labels based on the majority class among the \( k \) nearest neighbors\footnote{Implemented from Scikit-learn’s \texttt{KNeighborsClassifier}: \url{https://scikit-learn.org/stable/modules/neighbors.html\#classification}}. KNN is computationally efficient and performs well with standardized MFCC features.

    \renewcommand{\thefootnote}{\textit{f}}

    \item \textbf{Decision Tree (DT):} A tree-based model that sequentially splits data based on feature values\footnote{Implemented from Scikit-learn’s \texttt{DecisionTreeClassifier}: \url{https://scikit-learn.org/stable/modules/tree.html}}. Decision trees provide interpretability, showing feature importance for scream detection.

    \renewcommand{\thefootnote}{\textit{g}}

    \item \textbf{Multi-Layer Perceptron (MLP):} An artificial neural network with a single hidden layer, using backpropagation for training\footnote{Implemented from Scikit-learn’s \texttt{MLPClassifier}: \url{https://scikit-learn.org/stable/modules/neural_networks_supervised.html}}. MLPs capture non-linear patterns in audio data, essential for distinguishing complex scream features.

    \renewcommand{\thefootnote}{\textit{h}}

    \item \textbf{XGBoost:} An optimized gradient-boosting algorithm designed for performance\footnote{Implemented from XGBoost’s \texttt{XGBClassifier}: \url{https://xgboost.readthedocs.io/en/latest/python/python_api.html}}. XGBoost enhances predictive power by sequentially building decision trees, ideal for feature-rich datasets.

    \renewcommand{\thefootnote}{\textit{i}}

    \item \textbf{LightGBM:} A gradient-boosting model using histogram-based learning\footnote{Implemented from LightGBM’s \texttt{LGBMClassifier}: \url{https://lightgbm.readthedocs.io/en/latest/}}. LightGBM is efficient with large-scale data, enabling faster training.

    \renewcommand{\thefootnote}{\textit{j}}

    \item \textbf{CatBoost:} A gradient-boosting model that naturally handles categorical features\footnote{Implemented from CatBoost’s \texttt{CatBoostClassifier}: \url{https://catboost.ai/en/docs/concepts/python-reference_catboostclassifier}}. Its symmetric tree structure helps manage overfitting, particularly beneficial for balancing scream and non-scream data.
\end{itemize}

\subsubsection{Model Training and Evaluation}
Each classifier was trained on 80\% of the dataset, with the remaining 20\% allocated for testing. Features were standardized, and batch processing was conducted using data loaders. We evaluated each model using accuracy (\(Acc\)), precision (\(P\)), recall (\(R\)), and F1-score (\(F1\)), defined as:
\[
Acc = \frac{TP + TN}{TP + TN + FP + FN}, \quad P = \frac{TP}{TP + FP}, \quad R = \frac{TP}{TP + FN}, \quad F1 = 2 \cdot \frac{P \cdot R}{P + R}
\]

\subsection{Proposed Model with Wav2Vec2, Enhanced ConvNet and LSTM}

This model architecture utilizes Wav2Vec2 for feature extraction, followed by an Enhanced ConvNet composed of Conv1D layers with batch normalization and ReLU activation functions. Each Conv1D layer uses a kernel size of 3 and stride of 1, with max-pooling applied after each layer to reduce the time dimension. Dropout layers with a rate of 0.3 are included in the fully connected layers to prevent overfitting. The bidirectional LSTM layer captures sequential dependencies, and a series of fully connected layers finalizes the classification process.

\renewcommand{\thefootnote}{\textit{k}}
\vspace{-15pt}
\begin{figure}[h!]
    \centering
    \includegraphics[width=0.8\textwidth]{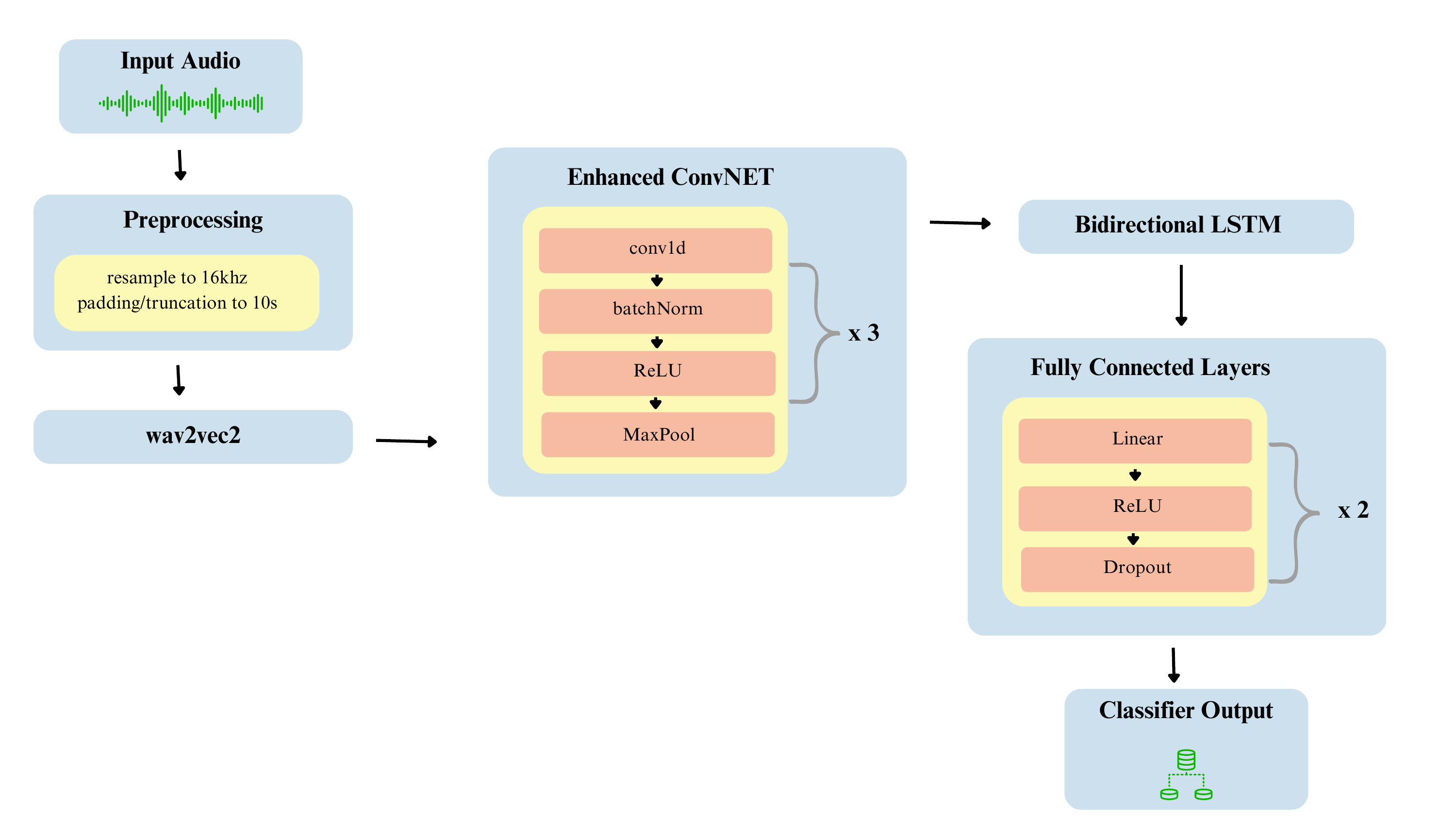}
    
    \caption{
    The system architecture of the DNN-based scream detection model, including Wav2Vec2 for feature extraction, an Enhanced ConvNet module with Conv1D layers, and a bidirectional LSTM layer. The architecture progresses from raw audio preprocessing through feature extraction, temporal modeling, and classification layers to output the detection result.
    }
    \label{fig:system_architecture}
\end{figure}
\vspace{-20pt}
\subsubsection{Data Preprocessing and Feature Extraction}

For the DNN-based scream detection model, audio samples were uniformly extended to 10 seconds, allowing for a consistent feature extraction process. Each audio sample was resampled to a 16 kHz rate (\(R = 16000 \, \text{Hz}\)) to match the optimal input requirements for the Wav2Vec2 model. Padding and truncation were applied to ensure that every sample reached the target length of \( T = R \times 10 \). This preprocessing enabled the DNN model to capture full audio segments while minimizing discrepancies caused by variable-length inputs.

Feature extraction utilized the pre-trained Wav2Vec2 model\footnote{Implemented using Hugging Face’s Transformers library: \url{https://huggingface.co/facebook/wav2vec2-base}}, which directly processes raw audio waveforms. Wav2Vec2 transforms the input waveform into high-dimensional embeddings, yielding hidden states that encapsulate intricate audio features, such as phonetic and temporal patterns. These hidden states formed the initial feature set \( X \in \mathbb{R}^{B \times T \times F} \), where:
- \( B \) represents the batch size,
- \( T \) is the sequence length, and
- \( F \) denotes the feature dimension, based on the Wav2Vec2 model's hidden layers.

The hidden states \( X \) were subsequently used as input to the Enhanced ConvNet-based classifier. This setup leveraged the Wav2Vec2 model’s capability to capture both high- and low-level sound attributes, providing a feature-rich representation conducive to scream detection.

\subsubsection{Model Architecture Components}

The proposed model consists of three main components: the EnhancedConvNet block, the LSTM layer, and the fully connected layers. Each component plays a crucial role in progressively transforming raw audio features into actionable outputs for scream detection. We have enhanced the convolutional component with advanced feature extraction techniques to capture intricate audio patterns, and we incorporated an LSTM layer to harness both spatial and sequential features in the audio data. Below is a detailed description of each component.

\paragraph{EnhancedConvNet Block}
The EnhancedConvNet component was structured to progressively learn higher-level representations through convolutional layers, batch normalization, ReLU activation, and max-pooling. Formally, each EnhancedConvNet layer can be expressed as:
\[
\text{EnhancedConvNet}(X) = \text{MaxPool}\left(\text{ReLU}\left(\text{BatchNorm}\left(\text{Conv1D}(X)\right)\right)\right)
\]
where:
- \( \text{Conv1D}(X) \) applies a 1D convolutional operation across the temporal dimension of \( X \), extracting localized features with a filter size of 3, a stride of 1, and padding of 1.
- \(\text{BatchNorm}\) normalizes the output of each convolutional layer to enhance stability during training.
- \(\text{ReLU}\) introduces non-linearity, allowing the model to learn complex, non-linear patterns in the audio data.
- \(\text{MaxPool}\) reduces the dimensionality by downsampling the output, thereby controlling the sequence length while retaining essential features.

This block was repeated with increasing channel sizes (from 64 to 256) across three stages, forming a robust spatial feature extractor for the scream and non-scream segments. The EnhancedConvNet configuration was designed to ensure minimal loss of critical temporal information during feature extraction.

\paragraph{LSTM Layer}
The LSTM layer operates on the output of the Enhanced ConvNet, capturing temporal dependencies in the high-level features. As a bidirectional LSTM, it processes the feature sequence in both forward and backward directions, effectively capturing contextual information:
\[
h_t = \tanh(W_{ih} x_t + b_{ih} + W_{hh} h_{(t-1)} + b_{hh})
\]
where:
- \( W_{ih} \) and \( W_{hh} \) are weight matrices for the input and hidden states, respectively.
- \( b_{ih} \) and \( b_{hh} \) are bias terms.
- \( x_t \) represents the input at time step \( t \), and \( h_t \) is the hidden state at the same step.

With a hidden size of 128 and bidirectional configuration, the LSTM outputs 256-dimensional embeddings for each time step, enabling the classifier to retain information from both past and future contexts within each sequence.

\paragraph{Fully Connected Layers}
The output from the LSTM layer is flattened and passed through a sequence of fully connected layers to produce class probabilities. The structure of these layers is as follows:
\[
\text{FC}(X) = \text{Dropout}\left(\text{ReLU}\left(\text{Linear}(X)\right)\right)
\]
Each fully connected layer is configured as:
1. A linear transformation layer that reduces dimensionality, followed by
2. ReLU activation to introduce non-linearity,
3. Dropout layers to prevent overfitting.

The final layer uses a softmax activation function to output class probabilities for scream and non-scream classifications.

\paragraph{Hyperparameters and Training}
The classifier was trained using AdamW as the optimizer, with two distinct learning rates:
- \( 1 \times 10^{-5} \) for the Wav2Vec2 parameters, ensuring fine-tuning on audio data,
- \( 1 \times 10^{-3} \) for the Enhanced ConvNet and LSTM components to expedite learning without destabilizing the base model. Class imbalance was addressed using weighted cross-entropy loss.

This layered architecture—leveraging Wav2Vec2 for feature extraction, Enhanced ConvNet for spatial representation, and LSTM for temporal dependencies—enables robust classification of scream and non-scream events in audio data.

\subsubsection{Training and Evaluation}
A weighted cross-entropy loss function was applied to handle class imbalance:
\[
\mathcal{L}_{CE} = -\frac{1}{N} \sum_{i=1}^{N} w_{c_i} \cdot \log(\hat{y}_i)
\]
where \( w_{c_i} \) denotes the weight for class \( c \), and \( \hat{y}_i \) is the predicted probability for sample \( i \). The AdamW optimizer was used with a learning rate of \(1 \times 10^{-5}\) for Wav2Vec2 and \(1 \times 10^{-3}\) for the Enhanced ConvNet classifier.

\subsection{Evaluation Metrics}
Both models were evaluated with accuracy, F1-score, and confusion matrices. For the DNN model, we also calculated the Equal Error Rate (EER) using the ROC curve:
\[
\text{EER} = \text{threshold} \left( \min_{\tau} | \text{FPR}(\tau) - (1 - \text{TPR}(\tau)) | \right)
\]
where \( \tau \) represents threshold values in the ROC analysis.

This detailed methodology outlines the unique setup of each classifier and the in-depth evaluation used to assess their performance for scream detection.

\subsection{Model Performance Analysis}

This paper presents a comparative analysis of various machine learning models tested on the scream detection dataset. The analysis includes Support Vector Classifier (SVC), Logistic Regression (LR), XGBoost (XGB), LightGBM (LGBM), CatBoost (CAT), Random Forest (RF), K-Nearest Neighbors (KNN), Decision Tree (DT), and Multi-Layer Perceptron (MLP) classifiers.

\begin{figure}[!ht]
    \centering
    \begin{minipage}{0.48\textwidth}
        \centering
        \begin{tikzpicture}
            \begin{axis}[
                width=\textwidth,
                height=0.75\textwidth,
                ybar,
                bar width=15pt,
                ylabel={Accuracy},
                xlabel={Models},
                ymin=0.75,
                ymax=0.92,
                symbolic x coords={SVC,LR,XGB,LGBM,CAT,RF,KNN,DT,MLP,Proposed Model},
                xtick=data,
                xticklabel style={rotate=45, anchor=east},
                nodes near coords,
                nodes near coords style={font=\tiny},
                title={Model Accuracy Comparison}
            ]
            \addplot coordinates {
                (SVC,0.77)
                (LR,0.79)
                (XGB,0.84)
                (LGBM,0.84)
                (CAT,0.85)
                (RF,0.84)
                (KNN,0.84)
                (DT,0.77)
                (MLP,0.82)
                (Proposed Model,0.91)
            };
            \end{axis}
        \end{tikzpicture}
        \caption{Overall Accuracy of Different Models.}
        \label{fig:accuracy}
    \end{minipage}
    \hfill
    \begin{minipage}{0.48\textwidth}
        \centering
        \begin{tikzpicture}
            \begin{axis}[
                width=0.85\textwidth,
                height=0.85\textwidth,
                xlabel={Predicted},
                ylabel={Actual},
                title={Confusion Matrix of Proposed Models Evaluation Set},
                xmin=-0.5,
                xmax=1.5,
                ymin=-0.5,
                ymax=1.5,
                xtick={0,1},
                ytick={0,1},
                xticklabels={Non-Scream,Scream},
                yticklabels={Non-Scream,Scream},
                ytick style={align=right},
                yticklabel style={rotate=90},
                colormap={blues}{
                    color(0)=(white)
                    color(1)=(rgb:blue,0.1)
                },
                colorbar,
                colorbar style={
                    ytick={0,200,400,600,700}
                },
                point meta min=0,
                point meta max=700,
                nodes near coords,
                nodes near coords style={font=\small},
                every node near coord/.append style={
                    text=black
                },
                enlargelimits=false,
                axis equal
            ]
            \addplot[matrix plot*,
                    point meta=explicit,
                    mesh/cols=2]
                coordinates {
                    (0,1) [776] (1,1) [45]
                    (0,0) [78]  (1,0) [426]
                };
            \end{axis}
        \end{tikzpicture}
        \caption{Confusion Matrix Results.}
        \label{fig:confusion}
    \end{minipage}
\end{figure}
\vspace{-20pt}
\begin{figure*}[!ht]
    \centering
    \begin{subfigure}[b]{0.48\textwidth}
        \begin{tikzpicture}
            \begin{axis}[
                width=\textwidth,
                height=0.75\textwidth,
                xlabel={Recall},
                ylabel={Precision},
                title={Precision vs Recall (Class 0: Non-Scream)},
                legend style={
                    at={(0.5,-0.3)},
                    anchor=north,
                    legend columns=5,
                    font=\tiny,
                    draw=black,
                    fill=white,
                    /tikz/column 2/.style={column sep=5pt},
                },
                xmin=0.8,
                xmax=0.97,
                ymin=0.75,
                ymax=0.93,
                grid=major
            ]
            \addplot[mark=*, mark size=2pt] coordinates {(0.92,0.76)}; 
            \addplot[mark=square*, mark size=2pt] coordinates {(0.90,0.78)}; 
            \addplot[mark=triangle*, mark size=2pt] coordinates {(0.89,0.86)}; 
            \addplot[mark=diamond*, mark size=2pt] coordinates {(0.90,0.85)}; 
            \addplot[mark=pentagon*, mark size=2pt] coordinates {(0.91,0.85)}; 
            \addplot[mark=+, mark size=3pt] coordinates {(0.91,0.84)}; 
            \addplot[mark=x, mark size=3pt] coordinates {(0.91,0.84)}; 
            \addplot[mark=star, mark size=3pt] coordinates {(0.82,0.81)}; 
            \addplot[mark=o, mark size=2pt] coordinates {(0.87,0.85)}; 
            \addplot[mark=asterisk, mark size=3pt] coordinates {(0.95,0.91)}; 
            \legend{SVC,LR,XGB,LGBM,CAT,RF,KNN,DT,MLP,Proposed Model}
            \end{axis}
        \end{tikzpicture}
        \caption{Precision-Recall Trade-off for Non-Scream Class.}
        \label{fig:pr_class0}
    \end{subfigure}
    \hfill
    \begin{subfigure}[b]{0.48\textwidth}
        \begin{tikzpicture}
            \begin{axis}[
                width=\textwidth,
                height=0.75\textwidth,
                xlabel={Recall},
                ylabel={Precision},
                title={Precision vs Recall (Class 1: Scream)},
                legend style={
                    at={(0.5,-0.3)},
                    anchor=north,
                    legend columns=5,
                    font=\tiny,
                    draw=black,
                    fill=white,
                    /tikz/column 2/.style={column sep=5pt},
                },
                xmin=0.5,
                xmax=0.97,
                ymin=0.65,
                ymax=0.93,
                grid=major
            ]
            \addplot[mark=*, mark size=2pt] coordinates {(0.53,0.80)}; 
            \addplot[mark=square*, mark size=2pt] coordinates {(0.59,0.79)}; 
            \addplot[mark=triangle*, mark size=2pt] coordinates {(0.76,0.81)}; 
            \addplot[mark=diamond*, mark size=2pt] coordinates {(0.74,0.82)}; 
            \addplot[mark=pentagon*, mark size=2pt] coordinates {(0.74,0.84)}; 
            \addplot[mark=+, mark size=3pt] coordinates {(0.73,0.83)}; 
            \addplot[mark=x, mark size=3pt] coordinates {(0.73,0.83)}; 
            \addplot[mark=star, mark size=3pt] coordinates {(0.69,0.70)}; 
            \addplot[mark=o, mark size=2pt] coordinates {(0.75,0.78)}; 
            \addplot[mark=asterisk, mark size=3pt] coordinates {(0.85,0.90)}; 
            \legend{SVC,LR,XGB,LGBM,CAT,RF,KNN,DT,MLP,Proposed Model}
            \end{axis}
        \end{tikzpicture}
        \caption{Precision-Recall Trade-off for Scream Class.}
        \label{fig:pr_class1}
    \end{subfigure}
    \caption{Precision-Recall Curves for Both Classes.}
    \label{fig:recallvsprecision}
\end{figure*}
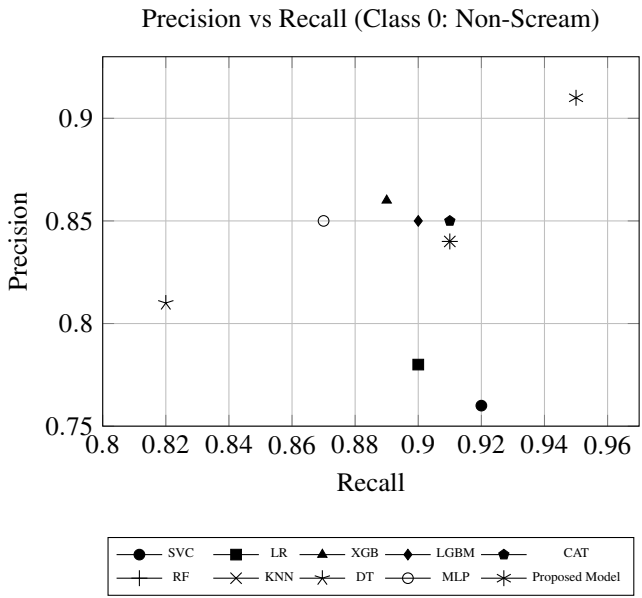
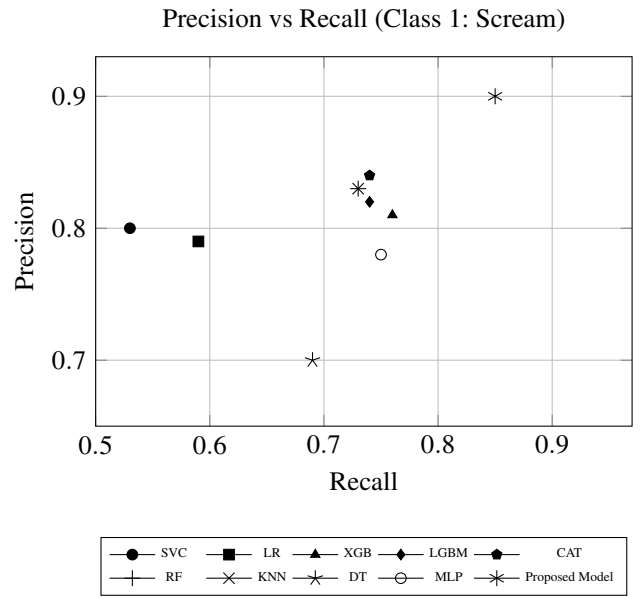

\begin{figure*}[!ht]
    \centering
    \begin{subfigure}[b]{0.60\textwidth}
        \begin{tikzpicture}
            \begin{axis}[
                width=\textwidth,
                height=0.5\textwidth,
                ybar=5pt,
                bar width=7pt,
                ylabel={F1-Score},
                xlabel={Models},
                ymin=0.6,
                ymax=0.95,
                symbolic x coords={SVC,LR,XGB,LGBM,CAT,RF,KNN,DT,MLP,Proposed Model},
                xtick=data,
                xticklabel style={rotate=45, anchor=east},
                legend style={
                    at={(0.5,-0.52)},
                    anchor=north,
                    legend columns=2,
                    column sep=1cm,
                    draw=black,
                    fill=white
                },
                nodes near coords style={font=\tiny},
                title={F1-Score Comparison by Class}
            ]
            \addplot coordinates {
                (SVC,0.83)
                (LR,0.84)
                (XGB,0.88)
                (LGBM,0.87)
                (CAT,0.88)
                (RF,0.88)
                (KNN,0.87)
                (DT,0.82)
                (MLP,0.86)
                (Proposed Model,0.93)
            };
            \addplot coordinates {
                (SVC,0.64)
                (LR,0.68)
                (XGB,0.79)
                (LGBM,0.78)
                (CAT,0.79)
                (RF,0.78)
                (KNN,0.77)
                (DT,0.70)
                (MLP,0.76)
                (Proposed Model,0.87)
            };
            \legend{Class 0 (Non-Scream), Class 1 (Scream)}
            \end{axis}
        \end{tikzpicture}
    \end{subfigure}
    \caption{F1-Score Analysis.}
    \label{fig:f1}
\end{figure*}
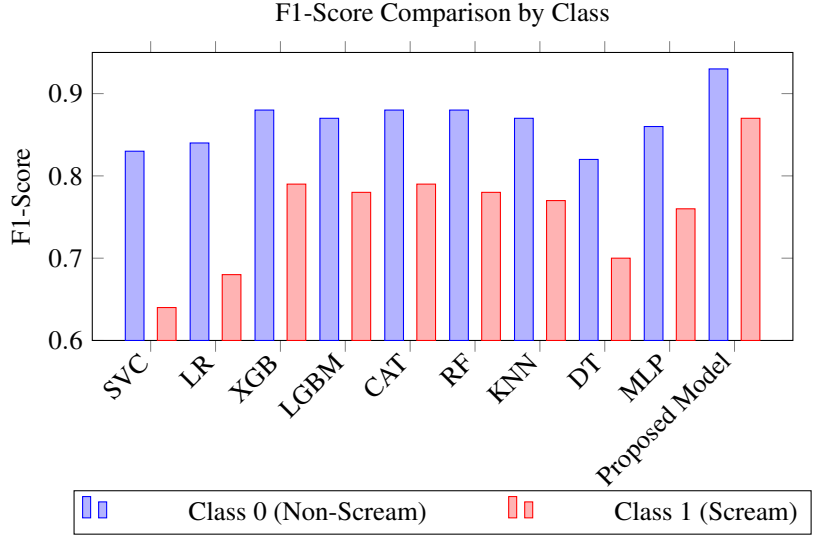
\vspace{1cm}
\subsection{Results Summary}

Our model demonstrates significant improvements over traditional classifiers in scream detection accuracy. Using Wav2Vec2 and Enhanced ConvNet for feature extraction and spatial analysis, our model achieved a 91\% accuracy rate—outperforming standard classifiers, which ranged from 77\% to 85\% accuracy. This combination of Wav2Vec2's phonetic and temporal feature extraction with Enhanced ConvNet and LSTM's spatial representation enables our model to accurately distinguish distress signals from typical construction noises, substantially reducing false positives.

As shown in Fig. \ref{fig:f1}, the F1 score for Non-Scream (Class 0) is notably higher than for Scream (Class 1), likely due to class imbalance in the dataset. Addressing this issue could be a valuable area for future research where researchers might expand the dataset with additional scream samples or employ augmentation techniques to improve class balance and enhance scream detection.
The results, depicted in Fig. \ref{fig:accuracy}
, \ref{fig:confusion}
, \ref{fig:recallvsprecision}
and \ref{fig:f1}
underscore the model's effectiveness and reliability. Although testing of this model in an actual scenario is yet to be done, these findings highlight the model’s potential as a robust solution for real-world, high-risk environments, marking a significant advancement in worker safety solutions.
\vspace{-5pt}
\section{Experimental Setup: Position Estimation}

Our proposed system integrates advanced acoustic signal processing with machine learning to create a robust framework for detecting and localizing distress signals in construction environments. The system operates through a carefully orchestrated sequence of processes, from initial audio capture to precise position estimation. Various TDOA methods have been proposed in literature, each with distinct advantages and limitations. Table \ref{tab:tdoa_comparison} summarizes these methods.

\begin{table}[h!]
    \centering
    \caption{Comparison of TDOA Methods in Construction Environments.}
        \label{tab:tdoa_comparison}  

    \begin{tabular}{p{2cm} p{3.5cm} p{3.5cm} p{3cm} p{3cm}}
        \toprule
        \textbf{Method} & \textbf{Advantages} & \textbf{Disadvantages} & \textbf{Best Use} & \textbf{Expression} \\
        \midrule
        Trilateration \cite{LI2020103309} & Simple setup, low compute, quick & Affected by machinery noise, poor with multiple equip., fails in enclosed spaces & Open sites, static sources & $\sqrt{(x - x_i)^2 + (y - y_i)^2} - \sqrt{(x - x_j)^2 + (y - y_j)^2} = c\Delta t_{ij}$ \\
        \midrule
        GCC-PHAT \cite{4538690} & Reflective and moving noise handling, adaptable, reliable & High-quality mics, high processing, freq. selection needed & Indoor construction, reflective surfaces & $\Psi_{PHAT}(f) = \frac{X_i(f) X_j^*(f)}{|X_i(f)||X_j(f)|}$ \\
                 & & & & $R_{PHAT}(\tau) = \int_{-\infty}^{\infty} \frac{X_i(\omega) X_j^*(\omega)}{|X_i(\omega)||X_j(\omega)|} e^{j \omega \tau} d\omega$ \\
        \midrule
        Basic GCC \cite{CHEN20114912} & Minimal processing, simple hardware & Poor with equipment noise, fails indoors & Outdoor prefab, single equipment & $R_{x_iy_j}(\tau) = \int_{-\infty}^{\infty} X_i(\omega) X_j^*(\omega)e^{j\omega\tau}d\omega$ \\
        \midrule
        MCCC \cite{Liu2022} & Good noise handling, tracks sources, high position accuracy & Complex setup, expensive & Large sites, multiple noise sources & $h_{DW}(\tau) = \prod_{i=1}^{N} \prod_{j=1}^{N} \rho_{ij}(\tau_i, \tau_j)$ \\
        \midrule
        SRP-PHAT \cite{app11010445} & High accuracy, full site coverage & Slow real-time use, costly & Safety monitoring, precise mapping & $\hat{P}_i(q) = \sum_{i=1}^{M} \sum_{j=i+1}^{M} \hat{R}_{ij}^{(i)} \left[ \tau_{i,j}(q) \right]$ \\
        \bottomrule
    \end{tabular}
\end{table}

\subsection{GCC-PHAT for TDOA Estimation in Complex Environments}
In reverberant and noisy construction environments, where signal reflections and environmental obstructions degrade accuracy, the Generalized Cross-Correlation with Phase Transform (GCC-PHAT) is a preferred technique for estimating Time Difference of Arrival (TDOA) \cite{4518172}, thus our research is centered around GCC-PHAT. The GCC-PHAT method computes the cross-correlation of signals received at different microphones by emphasizing the phase information, making it robust against reverberation.

The GCC-PHAT cross-correlation function \( R_{\text{PHAT}}(\tau) \) is defined as:
\begin{equation}
    R_{\text{PHAT}}(\tau) = \int_{-\infty}^{\infty} \frac{X_i(\omega) X_j^*(\omega)}{|X_i(\omega)||X_j(\omega)|} e^{j \omega \tau} d\omega
    \label{eq:gccphat}
\end{equation}
where \( X_i(\omega) \) and \( X_j(\omega) \) represent the Fourier transforms of the signals received at microphones \( i \) and \( j \), respectively, and \( \omega \) denotes the angular frequency. This cross-correlation function enhances the performance of TDOA estimation in complex construction environments by reducing the impact of multipath reflections and noise.

\subsection{TDOA Calculation and Position Estimation}
Given the TDOA estimates from the GCC-PHAT method, the sound source position can be estimated by optimizing a loss function that minimizes the discrepancy between the calculated TDOA values and those expected from the geometry of the microphone array. In an ideal setting, the TDOA between microphones \( i \) and \( j \) is given by:
\begin{equation}
    \tau_{ij} = \frac{d_i - d_j}{c}
    \label{eq:tdoa}
\end{equation}
where \( d_i \) and \( d_j \) are the distances from the sound source to microphones \( i \) and \( j \), respectively, and \( c \) is the speed of sound.

In complex construction environments with potential obstacles and reflections, position estimation can be improved by adjusting the estimated position to minimize the following loss function:
\begin{equation}
    \text{Loss} = \sum_{i,j} w_{ij} \left( \tau_{ij} - \frac{d_i - d_j}{c} \right)^2
    \label{weighted_loss}
\end{equation}

\subsection{Gradient Descent Optimization for Position Estimation}
To minimize the loss function defined in Equation \ref{weighted_loss}, gradient descent optimization is employed. At each iteration, a step in the direction of the negative gradient of the loss function is taken with a suitable learning rate to move towards the direction of the source \cite{6129502}.

\subsubsection{Gradient Descent Formulation}
Gradient descent iteratively updates the estimated source position \( (x, y, z) \) according to the rule:
\begin{equation}
    \mathbf{p}_{\text{new}} = \mathbf{p}_{\text{old}} - \eta \nabla \text{Loss}
    \label{eq:gradient_descent}
\end{equation}
where \( \eta \) is the learning rate, controlling the step size, and \( \nabla \text{Loss} \) represents the gradient of the loss function with respect to \( \mathbf{p} = (x, y, z) \).

The gradient for each component of \( \mathbf{p} \) is derived as:
\begin{equation}
    \frac{\partial \text{Loss}}{\partial x} = -2 \sum_{i,j} \left( \tau_{ij} - \frac{d_i - d_j}{c} \right) \frac{\partial}{\partial x} \left( \frac{d_i - d_j}{c} \right)
\end{equation}
and similarly for \( \frac{\partial \text{Loss}}{\partial y} \) and \( \frac{\partial \text{Loss}}{\partial z} \). These gradients are used to iteratively update the position estimate until convergence.

\[
\frac{\partial \text{Loss}}{\partial x} = -2 \sum_{i,j} \left( \tau_{ij} - \frac{d_i - d_j}{c} \right) \cdot \frac{1}{c} \left( \frac{\partial d_i}{\partial x} - \frac{\partial d_j}{\partial x} \right)
\]

\subsubsection{Advantages Over Linearization Techniques}
Gradient descent is especially suitable for scenarios involving non-linear loss functions, as found in reverberant construction environments where linear assumptions do not hold. While matrix trilateration and linearization methods can approximate source positions under ideal conditions, they struggle with the inaccuracies introduced by reverberations and noise. Gradient descent, however, allows for non-linear refinement of the position estimate, iteratively minimizing errors by directly addressing the discrepancies in TDOA measurements.

Talking about learning rates, gradient descent can achieve faster convergence and better adaptability to complex acoustic environments, ultimately yielding more accurate positioning compared to linearized techniques.

\vspace{-20pt}
\vspace{-20pt}
\vspace{-20pt}
\vspace{-20pt}
\vspace{-1cm}
\section{Conclusion and Discussion}\label{sec3}

The integration of real-time scream detection and localization marks a significant advancement in enhancing safety protocols on construction sites, especially in settings where traditional wearable and GPS-based systems may be insufficient. By focusing on audio detection, our system effectively addresses the physical and environmental limitations associated with wearables and GPS in complex environments. Leveraging Wav2Vec2 with an Enhanced ConvNet, the system achieves high accuracy in detecting screams even amidst substantial ambient noise.

Nevertheless, challenges persist in adapting the system to manage extreme noise levels and unexpected sound patterns that could affect detection specificity. Striking an optimal balance between sensitivity and specificity remains crucial to limiting false positives while ensuring swift responses during emergencies. Our source localization approach, which combines GCC-PHAT with an iterative optimization method, proved effective in achieving accurate Time Difference of Arrival (TDOA) estimation with computational efficiency. While the application of gradient descent improves accuracy, performance can still be influenced by environmental factors. The system’s high detection accuracy and minimal false positives suggest strong potential for adaptation in other high-risk industries, though ethical considerations around privacy remain, warranting safeguards like data encryption and scheduled purging.

In conclusion, our audio-based system offers promising advancements in worker safety, setting a foundation for further improvements in scalability, noise resilience, and privacy measures, which are essential for broader occupational applications. This work provides a basis for future development of intelligent, audio-focused safety solutions that could transform emergency response practices across various sectors globally.

\end{document}